\documentstyle[11pt,fleqn]{article}
\begin{document}
\thispagestyle{empty}
\date{}
\title{\bf COBE vs Cosmic Strings: An Analytical Model}
\author{Leandros Perivolaropoulos\thanks{Division of Theoretical Astrophysics,
Harvard-Smithsonian Center for Astrophysics
60 Garden St.
Cambridge, Mass. 02138, USA.}
\thanks{also Visiting Scientist, Department of Physics
Brown University Providence, R.I. 02912, U.S.A.}
}
\maketitle

\begin{abstract}
We construct a simple analytical model to study the effects of cosmic strings
on the
microwave background radiation. Our model is based on counting random multiple
impulses inflicted on photon trajectories by the string network between the
time
of recombination and today. We construct the temperature auto-correlation
function
and use it to obtain the effective power spectrum
index n, the rms-quadrupole-normalized amplitude  $Q_{rms-PS}$ and the
rms temperature variation smoothed on small angular scales. For the values of
the scaling solution parameters obtained in Refs.\cite{bb90},\cite{as90}
we obtain $n=1.14 \pm 0.5$, $Q_{rms-PS}=(4.5\pm
1.5) G\mu$ and $({{\Delta T}\over T})_{rms}=5.5 G\mu$. Demanding consistency of
these results with the COBE data leads to $G\mu=(1.7 \pm 0.7)\times 10^{-6}$
(where $\mu$ is the string mass per unit length), in good agreement with direct
normalizations of $\mu$ from observations.

\end{abstract}

\section{\bf Introduction}

\par
The recent detection of anisotropy\cite{cb92} in the Microwave Background
Radiation (MBR) by the COBE (COsmic Background Explorer) collaboration
has provided a new powerful experimental probe for testing theoretical
cosmological models. The DMR (Differential Microwave Radiometer)
instrument of COBE has provided temperature sky maps leading to
the rms sky variation $\sqrt{<({{\Delta T}\over T})^2>}$ (where
$\Delta T\equiv T(\theta_1)-T(\theta_2)$, and $\theta_1-\theta_2=60^\circ$ is
the beam separation in the COBE experiment)
and the rms quadrupole amplitude. A fit of the data
to spherical harmonic expansions has also provided
the angular
temperature auto-correlation function $C(\Delta\theta)\equiv  <{{\delta T}\over
T}(\theta)  {{\delta T}\over T}
(\theta ^\prime)>$ where $<>$ denotes averaging over all directions in the sky,
$\delta T(\theta)\equiv T(\theta)-<T>$ and $\Delta \theta=\theta
-\theta^\prime$.
This result was then used to obtain
the rms-quadrupole-normalized amplitude
$Q_{rms-PS}$ and the index $n$ of the power law fluctuation spectrum
assumed to be of the form $P(k) \sim k^n$. The published results are:
$$
({{\Delta T}\over T})_{rms}=(1.1 \pm 0.2)\times 10^{-5}
\eqno(1)
$$
$$
Q_{rms-PS}=(5.96\pm 0.75)\times 10^{-6}
\eqno(2)
$$
$$
n=1.1 \pm 0.5
\eqno(3)
$$
These results have imposed severe constraints on several cosmological models
for large scale structure formation. Even the standard CDM model with bias
$1.5\leq b_8 \leq 2.5$ is inconsistent with the COBE results for $H_0 > 50
km/(sec\cdot Mpc)$ and is barely consistent for $H_0\simeq 50 km/(sec\cdot
Mpc)$
\cite{cbt92}
\cite{be87}(Tensor mode perturbations have recently been shown however,
to make standard CDM with specific inflationary models
consistent with COBE for a wider cosmological
parameter region\cite{dhst92}).

It is therefore interesting to investigate the consistency of alternative
models with respect to the COBE measurments. The natural alternative to models
based on adiabatic Gaussian perturbations generated during inflation are models
where the primordial perturbations are created by topological defects like
cosmic strings global monopoles or textures.

Cosmological models based on topological defects have a single free parameter
$v$, the scale of symmetry breaking leading to defect formation. Consistency
with large scale structure observations, constrains this parameter to $G v^2
\simeq 10^{-6}$\cite{tb86}\cite{s86}
where G is Newton's constant. In what follows we will concentrate on the
case of cosmic strings.

Previous analytical studies of MBR anisotropies due to cosmic
strings\cite{ttb86} were based on the old picture of the cosmic string network
evolution\cite{at85} and therefore focused on the effects of cosmic string
loops. Loops however were later  shown by more detailed
simulations\cite{as90}\cite{bb90} to be unimportant compared to long strings.

More recently, numerical simulations have been used to investigate the MBR
predictions of cosmic string models\cite{bbs88} and comparison of these
predictions has
been made with the COBE data\cite{bsb92}. It was found that
for $G v^2 \simeq 10^{-6}$, ($v^2=\mu$ where $\mu$ is the mass
per unit length of the string) cosmic strings are
consistent with the COBE data for a wider range of cosmological parameters than
the standard CDM model. The numerical analysis that has led to this result
however, is rather complicated and
currently there is still some controversy among the
different groups\cite{as90}\cite{bb90}\cite{at89} about the details of the
simulations involved in the analysis.
In addition, studies based on string simulations have necessarily fixed
scaling solution parameters and therefore, the dependence of the results on
these
parameters can not be  revealed.
These arguments make the construction of an analytical model for the study of
the
effects of topological defects on the MBR, a particularly interesting prospect.
It is such an analytical model that we are constructing in this paper.

In particular, we use a multiple
impulse\cite{v92} approximation to obtain the temperature auto-correlation
function $C(\Delta \theta)$ predicted by the string model. From $C(\Delta
\theta)$ we obtain
the mass
per unit length $\mu$ of strings consistent with COBE and the effective power
spectrum index n predicted by the string model.

Our basic assumptions and approximations are the following:
\begin{enumerate}
\item We approximate the photon path from $t_{rec}$ to $t_0$ by a discrete set
of $N$ Hubble time-steps $t_i$ such that $t_{i+1}=2 t_i$. For $z_{rec}\simeq
1400$
we have $N\simeq \log_2[(1400)^{3/2}]\simeq 16$.

\item At each Hubble time-step the photon beam is affected only by the long
strings within a horizon distance from the beam. The effects of further strings
are cancelled due to the compensating scalar field radiation.

\item The combined effects of all strings present within a horizon distance of
the photon beam is a linear superposition of the individual effects.\cite{g91}

\item Each string that affects the photon beam induces a temperature variation
of the form\cite{ks84}\cite{s88}:
$$
{{\delta T}\over T}   =  4\pi G\mu \beta
$$
with
$$
\beta={\hat k}\cdot({\vec v_s}\gamma_s \times {\hat s})
$$
where $\hat{k}$, $\hat{s}$ and $\vec{v}_s$ are the unit wave-vector, the unit
vector along the string and the string velocity vector respectively (see Fig.
1).

\item The long strings within each horizon volume have random velocities
positions and orientations.

\item The effects of loops are unimportant compared to the effects of long
strings\cite{as90}\cite{bb90}.

\item Initial temperature inhomogeneities at the last scattering surface are
assumed negligible compared to those induced by the string network at later
times.

\end{enumerate}

 We will also use the result that for any function
$\theta_1 (\theta)$ that takes
random values as the independent variable varies we have:
$$
<f(\theta_1 (\theta))>_\theta=<f(\theta_1)>_{\theta_1}
$$
where $<>_\alpha$ implies averaging with respect to $\alpha$. We will call this
for obvious reasons the `ergodic hypothesis'.

Some of these assumptions are similar to those made in
corresponding numerical studies. Others (like assumption 2 which is an attempt
to take into account compensation) are improvements over those of the numerical
analyses. We
comment on the possibility of further improving some of these assumptions in
section 4.

\section{\bf The Temperature Correlation Function}
\par
We begin with a description of our model.
Consider a photon beam emerging from the last scattering surface at
$t_1=t_{rec}$. This beam of fixed temperature will initially suffer $M$ `kicks'
from the $M$ long strings within the horizon at $t_1$. At the Hubble time
$t_2 \sim 2 t_{rec}$, M strings, uncorrelated with the previous ones will be
within the horizon giving the photon M further `kicks' and the process will
continue
until the $N\simeq 16$  Hubble time-step corresponding to the present time
$t_0$ (
$t_0\simeq 2^{16} t_{rec}$). Therefore the total temperature shift in the
direction
$\theta$ (in what follows we consider fixed $\varphi$ and omit it when defining
direction unless otherwise needed) may be written as:
$$ {{\delta T}\over T}
(\theta)=4\pi G \mu \sum_{n=1}^N \sum_{m=1}^M \beta^{mn} (\theta)
\eqno(4)
$$
where $\beta^{mn}(\theta)$ corresponds to the mth string at the nth Hubble
time-step. Taking $\hat{v}^{mn}_s$ and $\hat{s}^{mn}$ to be unit vectors along
the string velocity and string length directions we may write
$\beta^{mn}(\theta)=v_s \gamma_s (\hat{k}(\theta)\cdot \hat{R}^{mn}_1(\theta))$
where $\hat{R}_1^{mn}(\theta)=\hat{v}^{mn}\times \hat{s}^{mn}$ is a unit vector
that varies randomly with m and n.

It is instructive for what follows to obtain ${{\delta T}\over T} (\theta)$
averaged over all directions. Defining $\xi\equiv 4\pi G\mu v_s \gamma_s$ we
have:
$$
<{{\delta T}\over T} (\theta)>=\xi \sum_{n,m}<\hat{k}(\theta)
\cdot \hat{R}_1^{mn}(\theta)>={\xi \over {4 \pi}}\sum_{m,n}\int d\cos
\theta d\varphi \cos \theta_1 (\theta,\varphi)
$$
where $\theta_1$ is the angle between $\hat{R}_1^{mn}(\theta)$ and
$\hat{k}(\theta)$
(see Fig. 3)
and since $\hat{R}_1^{mn}(\theta)$ varies randomly with $m,n$ and with
$\theta,\varphi$  on angular scales larger than $\theta_{rec}$, we may use the
ergodic hypothesis to obtain: $$
<{{\delta T}\over T} (\theta)>={\xi \over {4 \pi}}\sum_{m,n}\int d\cos
\theta_1 d\varphi_1 \cos \theta_1=0
$$
which is also the naively expected result.The same result could have been
obtained
without using the ergodic hypothesis by simply taking the large $M\times N$
limit and performing the sum over m,n before the $\theta, \varphi$ integration.

 Our goal is to investigate correlations of fluctuations :${{\delta T}\over
T}(\theta)  {{\delta T}\over T} (\theta ^\prime)$. Using (4) we have:
$$
{{\delta T}\over
T}(\theta) {{\delta T}\over T} (\theta ^\prime)=(4\pi G\mu)^2 \sum_{n,
n^\prime}^N \sum_{m,m^\prime}^M \beta^{mn}(\theta) \beta^{m^\prime
n^\prime}(\theta^\prime)
\eqno(5)
$$
Define now $\Delta \theta \equiv \theta-\theta ^\prime$ and focus on the case
when $\Delta \theta=\theta_p$ where $\theta_p$ is the angular size of the
horizon at the Hubble time-step $t_p$ ($1\leq p \leq 16$). Long strings present
at
time $t$ will inflict `kicks' on two photon beams separated by
$\Delta \theta=\theta_p$ that are uncorrelated for $t\leq t_p$ but are equal
for
$t>t_p$  when $\Delta \theta$ is within the horizon scale $t$ (see Fig. 2).
Therefore
$$
\hat{R}^{jk}_1 (\theta)=\hat{R}^{j^\prime k^\prime}_1 (\theta ^\prime)
\hskip .2cm {\rm iff}
\hskip .2cm
k>p,\hskip .1cm j=j^\prime, \hskip .1cm k=k^\prime
$$
$$
\hat{R}^{jk}_1 (\theta)\neq \hat{R}^{j^\prime k^\prime}_1 (\theta ^\prime)
\hskip .4cm {\rm otherwise}
$$
where $\neq$ here means `not equal and also uncorrelated'.
We may therefore split the sum (5) in two parts consisting of correlated and
uncorrelated products respectively.
$$
{{\delta T}\over
T}(\theta)  {{\delta T}\over T} (\theta ^\prime)=\xi^2[\sum_{n=n^\prime=p}^N
\sum_{m=m^\prime=1}^M (\hat{k}(\theta)\cdot \hat{R}_1^{mn}(\theta))
(\hat{k}(\theta^\prime)\cdot \hat{R}_1^{m^\prime n^\prime}(\theta^\prime))+
$$
$$
+\sum_{n,n^\prime}
\sum_{m,m^\prime} (\hat{k}(\theta)\cdot \hat{R}_1^{mn}(\theta))
(\hat{k}(\theta^\prime)\cdot \hat{R}_1^{m^\prime
n^\prime}(\theta^\prime))]\equiv \xi^2 [\Sigma_1+\Sigma_2]
$$
where $\Sigma_1$ refers to the terms of the sum that are correlated on a scale
$\Delta \theta$ while $\Sigma_2$ refers to the uncorrelated terms.
Averaging $\Sigma_1$ over all directions we obtain
$$
<\Sigma_1(\Delta \theta)>={1\over {4\pi}}\sum_{\theta_1^{mn}}\int d\cos\theta
d\varphi \cos\theta_1^{mn} (\theta,\varphi) \cos(\theta_1^{mn} (\theta,
\varphi)+\Delta
\theta)
$$
where $\theta_1^{mn}$ is the angle between $\hat{k}(\theta)$ and
$\hat{R}_1^{mn}(\theta)$ (see Fig.3). Since by the construction of the model
$\theta_1$ varies randomly from one correlated patch to another we may use the
ergodic hypothesis and replace the average over all directions with an average
over $\theta_1$. It is then easy to see that
$$
<\Sigma_1(\Delta \theta)>={{\cos(\Delta \theta)}\over 3} N_{cor}(\Delta \theta)
$$
where $N_{cor}(\Delta \theta)=M (N-p(\Delta \theta))$ is the number of terms
in $\Sigma_1$. For $\Sigma_2$ we have:
$$
<\Sigma_2(\Delta \theta)>={1\over {4\pi}}\int d\cos\theta
d\varphi\sum_{\theta_1^{mn}}\cos\theta_1^{mn}\sum_{\theta_2^{m^\prime
n^\prime}}
\cos(\theta_2^{m^\prime n^\prime}+\Delta\theta)=0
$$
Therefore we may write
$$
<{{\delta T}\over
T}(\theta)  {{\delta T}\over T} (\theta ^\prime)>={\xi^2 \over 3}
N_{cor}(\Delta
\theta) \cos(\Delta \theta)
\eqno(6)
$$
 Using the relations $t_p=2^p t_{rec}$ and $\Delta
\theta=\theta_{t_p}=z_p^{-1/2}$ (for $\Omega_\circ =1$)
where $z_p$ is the redshift
at $t_p$, it is straightforward to show that
$$
p(\Delta \theta)=3 \log_2({{\Delta \theta}\over \theta_{rec}})
\eqno(7)
$$
Using (6) and (7) we obtain
$$
C(\Delta \theta)\equiv <{{\delta T}\over
T}(\theta)  {{\delta T}\over T} (\theta ^\prime)>={\xi^2 \over 3} M
(N-3 \log_2({{\Delta \theta}\over \theta_{rec}})) \cos(\Delta \theta)
$$
Our assumptions for the construction of $C(\Delta \theta)$ clearly break down
for
$\Delta \theta<\theta_{rec}\simeq 1^\circ$ (since we must have $t_p\geq
t_{rec}$) and for $\Delta \theta > {{2 \pi}\over 9}$ (since we can not have
$N_{cor}<0$). However, the region of validity of (8) may be easily
extrapolated by slightly shifting $\Delta \theta$ by $\theta_{rec}\simeq
1^\circ$
to $\Delta \theta + \theta_{rec}$ in the log (thus extrapolating
$C(\Delta \theta)$ to $\Delta \theta=0$) and keeping $C(\Delta \theta)=0$ for
$\Delta \theta >{{2 \pi}\over 9}$ as is physically expected since $N_{cor}$
goes
to 0 at large angular separations. We may now write the extrapolated
auto-correlation function as: \newpage
$$
C(\Delta \theta)={\xi^2 \over 3} M
(N-3 \log_2(1+{{\Delta \theta}\over \theta_{rec}})) \cos(\Delta \theta)
\hskip .4cm 0\leq \Delta \theta \leq {{2\pi}\over 9}
$$
$$
\eqno(8)
$$
$$
C(\Delta \theta)=0 \hskip 1cm  {{2\pi}\over 9}\leq \Delta \theta \leq \pi
$$
The validity of this result for $C(\Delta\theta)$ has also been
verified\cite{pc1} by using the model introduced here to calculate directly the
rms sky variation and showing that the result is identical to the one obtained
below from (8) and (14) (see section 3).
In what follows we will use (8) to compare COBE's data
with the string model predictions. A typical value for M, the number of long
strings per horizon volume, is $M=10$ while for $z_{rec}=1400$ we have
$N\simeq 16$.

\section {Predictions of the Model}
\par
Consider an expansion of the temperature pattern on the celestial sphere in
spherical harmonics
$$
{{\delta T}\over T}=\sum_{lm} a_l^m Y_l^m (\theta, \varphi)
$$
Defining $C_l\equiv <\vert a_l^m\vert^2>$ it may be shown using the addition
theorem that
$$
C(\Delta\theta)\equiv <{{\delta T}\over T}(\theta) {{\delta T}\over
T}(\theta^\prime)>={1\over {4\pi}}\sum_l (2 l+1) C_l P_l(\cos
(\Delta \theta) $$
Assuming now a power spectrum of perturbations of the form
$P(k)\sim k^n$ it may be shown\cite{be87}
that
$$
C_l=C_2 {{\Gamma(l+{{n-1}\over 2}) \Gamma({{9-n}\over 2})}\over
{\Gamma(l+{{5-n}\over 2}) \Gamma({{3+n}\over 2})}}
\eqno(9)
$$
where $C_2\equiv {4\pi \over 5} (Q_{rms-PS})^2$, $Q_{rms-PS}$ being the
rms-quadrupole-normalized amplitude\cite{cb92}.

Since our model predicts $C(\Delta\theta)$, $C_l$ may be obtained for any $l$
using the orthogonality relations for the Legendre functions.
We may then find
$\bar{n}$ and $\bar{Q}_{rms-PS}$ that give the best fit to (9).
Using the symbol-manipulating package {\it Mathematica}\cite{w92}
we calculated
$C_l$ for $l=3...30$. We excluded the quadrupole $C_2$ from our fit
(as was done with the COBE data) and considered harmonics up to
$l_{max}=30$ to account for the small angle cutoff of COBE due to the finite
antena beam size. Our results are rather insensitive to the specific cutoff
$l_{max}$ for $l_{max}>15$.

Minimizing the sum
$$
\sum_{l=3}^{30} (C_l-C_2 {{\Gamma(l+{{n-1}\over 2}) \Gamma({{9-n}\over
2})}\over
{\Gamma(l+{{5-n}\over 2}) \Gamma({{3+n}\over 2})}})^2
\eqno(10)
$$
with respect to $C_2$ and $n$ we obtain:
$$
{\bar{n}}=1.14 \pm 0.24
\eqno(11)
$$
$$
{\bar{Q}_{rms-PS}}\equiv \sqrt{{5\over {4 \pi}}\bar{C}_2}=(4.5\pm 1.5) (G\mu)_6
(v_s \gamma_s)_{0.15} \sqrt{M_{10}}\times 10^{-6}
\eqno(12)
$$
as the parameters giving the best fit, where $(G\mu)_6$, $(v_s\gamma_s)_{0.15}$
and $M_{10}$ are the corresponding quantities in units of $10^{-6}$, $0.15$ and
$10$ respectively.
  The error bars were
$\sqrt{<(n_l-\bar{n})^2>_l}$ and similarly for $C_2$, where $n_l$(or $C_2^l$)
is obtained by equating to 0 the $l$th term of the sum (10) while fixing
$C_2$(or
$n$) to its best fit value.

In Fig. 4 we show a plot of $C_l$ vs $l$ obtained from $C(\Delta \theta)$ of
(8)
(continous thin line),
superimposed
with $C_l$ obtained from (9) for the best fit values $n=\bar{n}$ and
$Q_{rms-PS}=\bar{Q}_{rms-PS}$ (dashed thick line).
The corresponding results coming from the COBE
data are given by (3) and (2)
The agreement between (11) and (3) is not too surprising since it is well known
that topological defects with a scaling solution naturally lead to a slightly
tilted scale invariant Harrisson-Zeldovich spectrum on large angular scales.
Clearly however, the confirmed prediction of (11) provides a good test showing
that our simple analytic model has fairly robust and realistic features.

Comparison of the observational result (2) with the prediction (12) leads to an
estimate of G$\mu$. According to the recent simulations of Ref.\cite{as90},
prefered values of the scaling solution parameters on the horizon scale are
$v_s \gamma_s \simeq 0.15$ and $M \simeq 10$. Using these values we obtain
$$
G\mu=(1.3 \pm 0.5)\times 10^{-6}
\eqno(13)
$$
This result is in good agreement with the result obtained numerically
in Ref.\cite{bsb92}.

Another way to obtain an estimate for G$\mu$ is to compare the measured
$({{\Delta T}\over T})_{rms}\equiv \sqrt{<({{T(\theta_1)-T(\theta_2)}\over
T})^2>}$ with the corresponding value predicted by our model.

The rms temperature variation may be expressed in terms of the temperature
auto-correlation function for a two beam experiment by the relation:
$$
({{\Delta T}\over T})_{rms}^2=2 (C(0)-C(\alpha))
\eqno(14)
$$
where $\alpha$ is the beam separation angle ($60^\circ$ for COBE).
Since however $({{\Delta T}\over T})_{rms}^2$ for COBE
was obtained with a $3.2^\circ$ Gaussian beam smoothing
we must introduce a similar smoothing
before we compare with the COBE result.
Thus, we first expand $C(\theta)$ in spherical
harmonics and then reconstruct it using the window function
$W(l)=e^{-{l(l+1)}\over {2(17.8)^2}}$ as was done with the COBE data for
the construction of $C(\Delta \theta)$. This leads to
$$
C(\Delta \theta)={1\over {4 \pi}}\sum_{l=3}^{30} (2 l+1) C_l P_l(\cos
(\Delta \theta)) W(l)^2 $$
We now use the reconstructed $C(\theta)$ in (14) to obtain
$$
({{\Delta T}\over T})_{rms}=\sqrt{2.4 \xi^2 M/3}=0.55 (G\mu)_6 (v_s
\gamma_s)_{0.15} \sqrt{M_{10}}
\times 10^{-5}
$$
As we did for the rms quadrupole variation we may compare the predicted
rms variation with the observational result (1) of COBE to obtain the value
of $G\mu$ that makes our model consistent with the COBE data.
For $(v_s\gamma_s)=0.15$ and $M=10$ the resulting value for $G\mu$ is:
$$
G\mu=(2.0 \pm 0.5)\times 10^{-6}
\eqno(15)
$$
consistent with our corresponding result from $Q_{rms-PS}$. An
estimate of $G\mu$ may be obtained by averaging (13) and (15)
and extending the error bars to cover both (13) and (15)
$$
G\mu=(1.7 \pm 0.7)\times 10^{-6}
\eqno(16)
$$
The interesting consequenses of this result will be
discused in the next section.

\section {Discussion}
The main result of this letter is that the COBE data constrain the single
free parameter $G\mu$ of the cosmic string model to be in the range given
by (16). This range is in good agreement with normalizations of $G\mu$
from galaxy\cite{tb86}\cite{s86} and large scale structure
formation\cite{vv91}\cite{pbs90}. Therefore the cosmic
string model for structure formation remains a viable model, consistent
with the COBE data.

Recent studies have used large scale structure observations to derive
the dependence of $G\mu$ on the the bias
factor $b_8$, defined such that the rms fluctuation in a sphere of radius
8$h^{-1}$ Mpc is $1/b_8$\cite{as92a}\cite{as92b}.
In Fig. 5 we plot the allowed range of $b_8$
for four different cosmic string models, demanding that
the value of $G\mu$ in \cite{as92a}\cite{as92b} is consistent with (16).
For comparison we
also plot the allowed range of $b_8$ for the standard CDM model, with
cosmological parameters chosen to give the most reasonable value for it
\cite{be87}\cite{cbt92}.
 Our estimate for the allowed range of the bias factor in the string
model is slightly lower than
that of Ref.\cite{bsb92} but is clearly consistent with it. Since a value of
$b_8$
in the range 1 to 3 is generally well motivated physically,
Fig. 5 suggests that both the string model and standard CDM (with $h=0.5$) are
consistent	 with the COBE data.

 Even though our model is based on fairly simple assumptions it has provided
results that are not only self-consistent but also in good agreement with much
more complicated numerical analyses. This fact indicates that our assumptions
even though simple are fairly realistic and economical. In spite of that, there
is certainly room for improvement. One interesting impovement that
could be implemented in a straightforward way is the consideration of the
effects of loops. Even though we do not expect loops to affect the shape of the
auto-correlation function on the angular scales considered here they may
introduce
an overall multiplicative factor slightly larger than one, thus reducing the
predicted value of $G\mu$ by the same factor. Another improvement that would
modify our results in a similar way is the consideration of string induced
perturbations on the last scattering surface by strings present in the final
stages of the radiation era. Finally, it would be interesting to include the
effects of compensation in a more realistic way. For example, instead of
introducing an abrupt cutoff of the deficit angle on the horizon scale we could
smoothly reduce it to zero as the horizon scale is aproached. The effect of
this
modification would increase the predicted value of $G\mu$ thus
tending to cancel the effect of the previous two modifications.

It is straightforward to apply the analysis presented here to other seed based
cosmological models including the global monopole and texture models. Work in
that direction is currently in progress.

\section {Acknowledgements}
\par
I wish to thank R. Brandenberger, Richhild Moesner and T. Vachaspati for
interesting discussions and for providing heplful comments after
reading the paper.
This work was supported by a CfA Postdoctoral Fellowship.

\vskip 1cm
\centerline {\Large \bf Figure Captions}
\vskip .5cm
{\bf Figure 1} : \hfil\break
The geometry of the vectors $\vec{v}$, $\hat{s}$ and $\hat{k}$.
\vskip .5cm
{\bf Figure 2} :\hfil\break
Two photon beam paths in directions $\theta$ and $\theta^\prime$ from $t_{rec}$
to $t_0$. The horizon in three Hubble time-steps is also shown. The effects
of strings during the first two Hubble time-steps are uncorrelated for
the particular angular beam separation shown.
\vskip .5cm
{\bf Figure 3} :\hfil\break
The geometry of the vectors $\hat{k}(\theta)$ and $\hat{R}_1^{mn}(\theta)$ in
the term $\Sigma_1(\Delta \theta)$.
\vskip .5cm
{\bf Figure 4} :\hfil\break
The multipole coefficients $C_l$ (continous thin line) superimposed with the
best fit of (9) (dashed thick line), obtained by using $n=1.14$.
\vskip .5cm
{\bf Figure 5} :\hfil\break
The allowed range of the bias factor $b_8$ for four cosmic string based models
obtained using our result for $G\mu$ and the results of
Ref. \cite{as92a}\cite{as92b}. The following cases are shown:\\
a) Strings + CDM, h=1 and h=0.5 \\
b) Strings + HDM, h=1 and h=0.5 \\
 The allowed range for standard CDM with $h=0.5$ and
$\Omega_b=0.1$ (the most consistent case with COBE's data) is also shown for
comparison.

\end{document}